\documentclass[aps,twocolumn,prb,floatfix,showpacs,superscriptaddress]{revtex4-1}

\usepackage{amsmath}
\usepackage{amssymb}
\usepackage{amsfonts}
\usepackage[pdftex]{graphicx}
\usepackage{units}
\usepackage[usenames]{color}
\usepackage{dcolumn}
\usepackage{bm}
\usepackage{float}
\usepackage[utf8]{inputenc}
\usepackage[colorlinks=true,citecolor=blue,linkcolor=blue]{hyperref}

\renewcommand{\vec}[1]{\ensuremath{\boldsymbol{#1}}}

\newcommand{\rdep}{\left(\vec{r}\right)}
\newcommand{\e}{\mathcal{E}}
\newcommand{\E}{\vec{\mathcal{E}}}
\newcommand{\jas}{j^{s}_{a}\rdep}
\newcommand{\jasop}{\hat{j}^{s}_{a}\rdep}
\renewcommand{\r}{\vec{r}}
\renewcommand{\k}{\vec{k}}
\newcommand{\vv}{\vec{v}}
\newcommand{\s}{\vec{s}}
\def\ket#1{\vert#1\rangle}
\def\bra#1{\langle#1\vert}
\def\ip#1#2{\langle#1\vert#2\rangle}
\def\me#1#2#3{\langle#1\vert#2\vert#3\rangle}
\DeclareMathOperator\Tr{Tr}
\def\wt#1{\widetilde{#1}}

\begin{document}

\title{Local spin Hall conductivity}

\author{Tom\'{a}\v{s} Rauch} 
\affiliation{Friedrich-Schiller-University Jena, 07743 Jena, Germany}

\author{Franziska Töpler} 
\affiliation{Martin-Luther-University Halle-Wittenberg, 06120 Halle/Saale, Germany}

\author{Ingrid Mertig} 
\affiliation{Martin-Luther-University Halle-Wittenberg, 06120 Halle/Saale, Germany}

\date{\today}

\begin{abstract}
 The spin Hall conductivity is usually calculated for periodic solids, while the knowledge of local contributions from inhomogeneous regions like interfaces or surfaces would be desirable. In this work we derive a local expression for the spin Hall conductivity of two-dimensional systems in analogy to the local anomalous Hall conductivity discussed recently. Our approach is applicable to heterogeneous fully bounded systems and nanoribbons with both open and periodic boundary conditions.
\end{abstract}

\maketitle

\section{Introduction}
Spintronics is a discpline which studies the electronic spin degree of freedom and aims at using its unique properties to invent functional devices, similar to electronics. One of its most prominent phenomena is the spin Hall effect, discovered in 1971\cite{Dyakonov1971a,Dyakonov1971b}, where an external electric field $\E$ induces a spin current $\vec{j}^{s}$ polarized along the $s$-direction and propagating in the direction perpendicular to $\E$, in analogy to the anomalous Hall effect, where a charge current is propagating instead of the spin current (in absence of external magnetic field). While the original discovery proposed an extrinsic mechanism, in 2004 an intrinisic spin Hall effect has been proposed~\cite{Sinova2004} and experimentally confirmed one year later~\cite{Wunderlich2005}, as a consequence of the spin-orbit coupling.

In the linear response regime, the strength of the spin Hall effect is described by the spin Hall conductivity (SHC) tensor, which can be derived from the Kubo-theory~\cite{Sinova2004} and calculated for periodic solids. Unfortunately, the assumption of a perfect periodic crystal is usually not valid for spintronics devices, whose surfaces, electronic contacts or internal inhomogenities can influence or even be responsible for its performance, as is the case in, e.g., the giant magnetoresistance effect~\cite{Baibich1988,Binasch1989}, a first important spintronics application.
Recently, spin currents and spin-orbit torques at heterostructure interfaces were investigated with a large effort~\cite{PhysRevB.94.104419,PhysRevB.94.104420,Baek2018}. The spin Hall effect has been identified as one of their possible sources.
Therefore, it is desirable to have an expression for the SHC, which would be valid locally and could describe, e.g., the contribution of a certain interface to the SHC of an entire heterostructure. 
Such approaches can be already found in the literature. For example, in Ref.~\onlinecite{PhysRevB.90.174423} the calculated spin-orbit torque is projected on muffin tin spheres surrounding different atoms to obtain local information and in Ref.~\onlinecite{PhysRevB.92.064415} the authors project the velocity-operator on the Wannier functions which are the basis of their model. These are rather simple approaches, because usually different local projections are possible and they are not necessarily equivalent~\cite{Rauch2018}. The authors of Ref.~\onlinecite{PhysRevLett.116.196602}, on the other hand, directly calculate the local electron and spin current density, which is well-defined. These quantities are then used to evaluate the interface spin Hall angle.

In our work we exploit once more the analogy between the spin and anomalous Hall effect expressed by the anomalous Hall conductivity (AHC). Recently, a local description of the AHC has been derived for both insulators~\cite{Bianco2011} and metals~\cite{Marrazzo2017} by transforming the standard Kubo-formula for periodic systems to real space, sometimes referred to as local (Chern) marker. As noted in Ref.~\onlinecite{Rauch2018}, this approach only considers the geometric part of the AHC, whereas the non-geometric contributions (vanishing for periodic systems) are omitted. Since the non-geometric contributions might be important in inhomogeneous systems, in this work we will adopt the approach of Ref.~\onlinecite{Rauch2018} to derive a local SHC (LSHC) from a local spin current density, which includes both the geometric and non-geometric parts.
In fact, our starting point is thus the same one as in Ref.~\onlinecite{PhysRevLett.116.196602}.
We carry out these derivations for different geometries with open and mixed (open and periodic) boundary conditions in Sec.~\ref{sec:LSHC} and we show that the LSHC of a bounded system transforms to the usual Kubo-formula for periodic systems after performing the thermodynamic limit. In Sec.~\ref{sec:numerical} we show the results of test calculations of the LSHC on the Kane-Mele model in both insulating and metallic regime. Finally, we summarize our work in Sec.~\ref{sec:conclusion}.

\section{Local spin Hall conductivity}
\label{sec:LSHC}
To derive an equation for the LSHC, we follow the ideas of Ref.~\onlinecite{Rauch2018} where the local AHC (LAHC) has been studied. The LSHC is defined as 
\begin{equation}
 \label{eq:sigma}
 \sigma^{s}_{ab}\rdep = \left. \frac{\partial \jas}{\partial \e_b } \right|_{\E=0}.
\end{equation}
$\E$ is a static homogeneous external electric field and $\jas$ is the induced spin current density. Since we are interested in the spin Hall effect, we will only consider the off-diagonal components of the spin-conductivity tensor with $a \neq b$. We use the conventional definition of the spin-current operator
\begin{equation}
    \label{eq:spin_cur_op_glob}
    \hat{j}^{s}_{a} = \frac{1}{2} \left(\hat{v}_a\hat{s}^{s} + \hat{s}^{s}\hat{v}_a\right)
\end{equation}
with the velocity operator $\hat{\vv} = (1/i\hbar)[\hat{\r},\hat{H}]$ and the spin operator $\hat{\s} = (\hbar/2) \hat{\vec{\sigma}}$, where $\hat{\vec{\sigma}}$ is the vector of the Pauli matrices. The local spin-current operator is then 
\begin{align}
    \label{eq:spin_cur_op}
    \jasop &= 1/2 \left(\ket{\r}\bra{\r}\hat{j}^{s}_{a}  + \hat{j}^{s}_{a} \ket{\r}\bra{\r}\right) \nonumber\\
           &= \frac{1}{2}\left[\ket{\r}\bra{\r} \frac{1}{2} \left(\hat{v}_a\hat{s}^{s} + \hat{s}^{s}\hat{v}_a\right) \right. + \nonumber\\ 
           & + \left. \frac{1}{2} \left(\hat{v}_a\hat{s}^{s} + \hat{s}^{s}\hat{v}_a\right) \ket{\r}\bra{\r} \right].
\end{align}
We obtain the local spin current density as
\begin{equation}
    \label{eq:spin_cur_den}
    \jas = \Tr\left[\hat{P}\jasop\right] = \Re\left[ \me{\r}{\frac{1}{2} \left(\hat{v}_a\hat{s}^{s} + \hat{s}^{s}\hat{v}_a\right)\hat{P}}{\r} \right],
\end{equation}
where $\hat{P}$ is the projection operator on the occupied states. We are aware of the existence of other definitions of the spin current in the literature~\cite{Vernes2007,Zhang2008}, which try to solve some problems of the conventional definition connected to the conservation of the spin current. Nevertheless, we work with the conventional definition (Eq.~\eqref{eq:spin_cur_den}), since starting from it, we obtain expressions for both the LSHC, as well as the common bulk SHC~\cite{Sinova2004}.

We continue by plugging Eq.~\eqref{eq:spin_cur_den} into Eq.~\eqref{eq:sigma}. Note that in the presence of electric field we can write the full Hamiltonian as $\hat{H} = \hat{H}_0 + e \E\cdot\r$ with the unperturbed Hamiltonian $\hat{H}_0$. Since $\left[\hat{r}_a,\hat{r}_b\right]=0$, the velocity operator is $\hat{\vv} = \hat{\vv}_0 = (1/i\hbar)[\hat{\r},\hat{H}_0]$ and the electric field modifies $\hat{P}$ only. We thus arrive at %
\begin{equation}
    \label{eq:lshc_fin}
    \sigma^{s}_{ab}\rdep = \Re\left[ \me{\r}{\frac{1}{2} \left(\hat{v}_a\hat{s}^{s} + \hat{s}^{s}\hat{v}_a\right)\frac{\partial\hat{P}}{\partial\e_b}}{\r} \right]
\end{equation}
where we obtained the partial derivative of the projection operator from perturbation theory as
\begin{equation}
    \label{eq:dP_dE}
    \frac{\partial\hat{P}}{\partial\e_b} = -e \sum_{v,c} \left( \ket{c}\frac{r_{b,cv}}{E_{cv}}\bra{v} + \ket{v}\frac{r_{b,vc}}{E_{cv}}\bra{c} \right).
\end{equation}
$\ket{v}$ and $\ket{c}$ denote the valence and conduction eigenstates of the unpertubed Hamiltonian with eigenenergies $E_v$ and $E_c$, respectively, $r_{b,cv} = \me{c}{\hat{r}_{b}}{v}$, and $E_{cv} = E_c - E_v$.

Eq.~\eqref{eq:lshc_fin} can be directly applied to calculate the LSHC of finite (bounded) systems with open boundary conditions (OBC) where $r_{b,cv}$ is always well-defined. Note that the derivation is completely analogous to the one of the LAHC in Ref.~\onlinecite{Rauch2018}.

\subsection{Periodic boundary conditions}
In the following we show that under the assumption of periodic boundary conditions (PBC), i.e., for an open system, we obtain the standard Kubo formula for the SHC of periodic systems. We assume a two-dimensional periodic system with a unit cell area $A_c$. The bulk SHC is obtained by averaging the LSHC in one unit cell,
\begin{equation}
    \label{eq:shc1}
    \sigma^{s}_{ab} = \frac{1}{A_c} \int dxdy\ \sigma^{s}_{ab}\rdep.
\end{equation}
We now plug Eq.~\eqref{eq:lshc_fin} into Eq.~\eqref{eq:shc1}. For this step we need the ground-state projection operator in terms of the occupied states of the periodic system,
\begin{equation}
    \label{eq:P_k}
    \hat{P}_0 = \frac{1}{N_{\k}}\sum_{\k,v} \ket{\psi_{\k,v}}\bra{\psi_{\k,v}},
\end{equation}
where the $N_{\k}$ wavevectors $\k = (k_x,k_y)$ cover the Brillouin zone. In analogy to Eq.~\eqref{eq:dP_dE} we obtain
\begin{equation}
    \label{eq:dP_dE_per}
    \frac{\partial\hat{P}}{\partial\e_b} = \frac{1}{N_{\k}} \sum_{\k,v} e^{i\k\cdot\hat{\r}} \left( \ket{\wt{\partial}_{\e_b} u_{\k,v}}\bra{u_{\k,v}} + \ket{u_{\k,v}}\bra{\wt{\partial}_{\e_b} u_{\k,v}} \right)e^{-i\k\cdot\hat{\r}}
\end{equation}
with $u_{\k,v}$ the cell-periodic part of $\psi_{\k,v}$ and 
\begin{equation}
    \label{eq:proj_der}
    \ket{\wt{\partial}_{\e_b} u_{\k,v}} = ie \sum_{c} \ket{u_{\k,c}} \frac{\hbar v_{b,\k,cv}}{E^{2}_{\k,cv}}
\end{equation}
the electric-field derivative $\ket{\partial_{\e_b} u_{\k,v}}$ projected on the empty states. We further defined $v_{b,\k,cv} = \me{u_{\k,c}}{\hat{v}_{b,\k}}{u_{\k,v}}$ and $\hat{v}_{b,\k} = e^{-i\k\cdot\hat{\r}} \hat{v}_b e^{i\k\cdot\hat{\r}}$.

Finally, we let $N_{\k}\rightarrow \infty$ and obtain
\begin{widetext}
\begin{equation}
    \label{eq:shc2}
    \sigma^{s}_{ab} = \frac{1}{4\pi^{2}}\int d^{2}k \int_{A_c} dxdy\ \sum_v \Re \left[ \bra{\r} \frac{1}{2}  \left(\hat{v}_a\hat{s}^{s} + \hat{s}^{s}\hat{v}_a\right) \left( \ket{\wt{\partial}_{\e_b} u_{\k,v}}\ip{u_{\k,v}}{\r} + \ket{u_{\k,v}}\ip{\wt{\partial}_{\e_b} u_{\k,v}}{\r} \right)  \right],
\end{equation}
\end{widetext}
again in full analogy to the LAHC in a slab geometry~\cite{Rauch2018}. Using Eq.~\eqref{eq:proj_der} and evaluating the real-space integral in Eq.~\eqref{eq:shc2} we arrive at
\begin{widetext}
\begin{equation}
    \label{eq:shc3}
    \sigma^{s}_{ab} = -\frac{e\hbar}{4\pi^{2}}\int d^{2}k\ \Im \sum_{v,c} \frac{j^{s}_{a,\k,vc}v_{b,\k,cv} - j^{s}_{a,\k,cv}v_{b,\k,vc}}{E^{2}_{\k,cv}} = -\frac{e\hbar}{2\pi^{2}}\int d^{2}k\ \Im \sum_{v,c} \frac{j^{s}_{a,\k,vc}v_{b,\k,cv}}{E^{2}_{\k,cv}}
\end{equation}
\end{widetext}
where we defined $j^{s}_{a,\k,vc} = \me{u_{\k,v}}{\frac{1}{2} \left(\hat{v}_a\hat{s}^{s} + \hat{s}^{s}\hat{v}_a\right)}{u_{\k,c}}$. Eq.~\eqref{eq:shc3} is the usual bulk formula for the SHC of periodic materials~\cite{Sinova2004}.

\subsection{Mixed boundary conditions}
Calculating LSHC from Eq.~\eqref{eq:lshc_fin} can be computationally demanding for large finite systems. Therefore, it might be useful to turn to the nanoribbon geometry with mixed boundary condition. We recall that in the symbol $\sigma^{s}_{ab}$ $a$ denotes the propagation direction of the spin current, $b$ denotes the orientation of the electric field, and $s$ is the polarization direction of the spin current. For the nanoribbon geometry, we introduce a new pair of indices $(\alpha,\beta) = (x,y)$, $\alpha \neq \beta$, where $\alpha$ and $\beta$ denote the direction with OBC (bounded) and PBC (open), respectively. The local information shall be kept in the OBC direction and we thus wish to calculate
\begin{align}
    \label{eq:sigma_nanor_gen}
    \sigma^{s}_{ab}(r_{\alpha}) &= \frac{1}{L_{\beta}} \int_{L_{\beta}} dr_{\beta}\ \sigma^{s}_{ab}(r_{\alpha},r_{\beta}) \nonumber \\
    &= \frac{1}{L_{\beta}} \int_{L_{\beta}} dr_{\beta}\ \Re\left[ \me{\r}{ \hat{j}^{s}_{a} \frac{\partial \hat{P}}{\partial \e_b} }{\r}\right]
\end{align}
with $L_{\beta}$ the length of the periodic unit cell along ${\beta}$ and the ground-state projector as in Eq.~\eqref{eq:P_k},
\begin{align}
    \label{eq:P_kb}
    \hat{P}_0 &= \frac{1}{N_{k_{\beta}}}\sum_{k_{\beta},v} \ket{\psi_{k_{\beta},v}}\bra{\psi_{k_{\beta},v}} \nonumber \\
    &= \frac{1}{N_{k_{\beta}}}\sum_{k_{\beta},v} e^{ik_{\beta}\hat{r}_{\beta}}\ket{u_{k_{\beta},v}}\bra{u_{k_{\beta},v}} e^{-ik_{\beta}\hat{r}_{\beta}}.
\end{align}

In the following it is necessary to distinguish between the two cases where the spin current is oriented along the direction with PBC and the electric field along the direction with OBC, and vice versa.

\subsubsection{Electric field along PBC}
When the electric field is oriented along the PBC direction, we have $a=\alpha$ and $b=\beta$. The spin-current operator is thus
\begin{equation}
    \label{eq:jas_a_eq_a}
    \hat{j}^{s}_{a} = \frac{1}{2i\hbar}\left( \left[\hat{r}_a,\hat{H}_0\right] \hat{s}^{s} + \hat{s}^{s} \left[\hat{r}_a,\hat{H}_0\right] \right)
\end{equation}
and
\begin{align}
    \label{eq:dP_dE_b_eq_b}
    \frac{\partial\hat{P}}{\partial\e_b} &= \frac{1}{N_{b}} \sum_{k_b,v} e^{ik_b\hat{r}_b} \left( \ket{\wt{\partial}_{\e_b} u_{k_b,v}}\bra{u_{k_b,v}} \right.+ \nonumber \\ 
    & + \left. \ket{u_{k_b,v}}\bra{\wt{\partial}_{\e_b} u_{k_b,v}} \right)e^{-ik_b\hat{r}_b}.
\end{align}
We thus obtain for the $r_a$-resolved LSHC
\begin{widetext}
\begin{equation}
    \label{eq:shc_a_eq_a}
    \sigma^{s}_{ab}(r_a) = \frac{e}{4\pi} \int dk_b \int_{L_b} dr_b\ \Re \sum_{v,c}  \bra{\r} \left( \left[\hat{r}_a,\hat{H}_0\right] \hat{s}^{s} + \hat{s}^{s} \left[\hat{r}_a,\hat{H}_0\right]  \right) \left( \ket{u_{k_b,c}} \frac{v_{b,k_b,cv}}{E^{2}_{k_b,cv}} \bra{u_{k_b,v}} + \ket{u_{k_b,v}} \frac{v_{b,k_b,vc}}{E^{2}_{k_b,cv}} \bra{u_{k_b,c}} \right) \ket{\r} .
\end{equation}
\end{widetext}

\subsubsection{Electric field along OBC}
The situation of the electric field being oriented along the OBC direction corresponds to $a = \beta$ and $b = \alpha$. We write the spin-current operator as in Eq.~\eqref{eq:spin_cur_op_glob} and
\begin{align}
    \label{eq:dP_dE_a_eq_b}
    \frac{\partial\hat{P}}{\partial\e_b} &= -\frac{e}{N_{a}} \sum_{k_a,v,c} e^{ik_a\hat{r}_a} \left( \ket{u_{k_a,c}} \frac{r_{b,k_a,cv}}{E_{k_a,cv}} \bra{u_{k_a,v}} \right.+ \nonumber \\ 
    & + \left. \ket{u_{k_a,v}}\frac{r_{b,k_a,vc}}{E_{k_a,cv}} \bra{u_{k_a,c}} \right)e^{-ik_a\hat{r}_a}.
\end{align}
The $r_b$-resolved LSHC for this geometry becomes
\begin{widetext}
\begin{equation}
    \label{eq:shc_a_eq_b}
    \sigma^{s}_{ab}(r_b) = -\frac{e}{4\pi} \int dk_a \int_{L_a} dr_a\ \Re \sum_{v,c}  \bra{\r} \left( \hat{v}_{a,k_a} \hat{s}^{s} + \hat{s}^{s} \hat{v}_{a,k_a} \right) \left( \ket{u_{k_a,c}} \frac{r_{b,k_a,cv}}{E_{k_a,cv}} \bra{u_{k_a,v}} + \ket{u_{k_a,v}} \frac{r_{b,k_a,vc}}{E_{k_a,cv}} \bra{u_{k_a,c}} \right) \ket{\r} .
\end{equation}
\end{widetext}

\section{Numerical calculations}
\label{sec:numerical}
To test the validity of our derivations, we perform numerical tight-binding (TB) calculations of the LSHC for both a finite system with open boundary conditions and a system in the nanoribbon geometry with mixed boundary conditions. 
As the main test of the validity of our local formulation
we show that in the central (bulk) region of a sufficiently large flake or nanoribbon the LSHC equals the bulk SHC calculated for a corresponding periodic system using the standard formula (Eq.~\eqref{eq:shc3}).
While this can be straightforwadly shown for insulators, a more complicated treatment will be necessary for metals.

For the TB calculations we used the PythTB package~\cite{PythTB}. In TB, the basis set are the localized TB orbitals $\phi_i \rdep = \ip{\r}{i}$. We thus have to replace $\ket{\r}$ with $\ket{i}$ and calculate a site-resolved LSHC $\sigma^{s}_{ab}(l) = \sum_{i\in l} \sigma^{s}_{ab}(i)$ where we sum over all TB orbitals $i$ located on the site $l$. The matrix elements of the position operator are calculated in TB using the on-site approximation $\me{i}{\hat{\r}}{j}=\vec{\tau}_i\delta_{ij}$ where $\vec{\tau}_i$ is the position of the orbital $i$~\cite{Boykin2010}.

\subsection{Kane-Mele model}
For the numerical test calculations we use the Kane-Mele model~\cite{Kane2005a} which is expressed by the Hamiltonian
\begin{align}
    \label{eq:KM_model}
    H &= E_p \sum_{i} \xi_i c^{\dagger}_{i,\gamma} c_{i,\gamma} + t \sum_{\langle ij\rangle} c^{\dagger}_{i,\gamma} c_{j,\gamma} + \nonumber \\
    & + i \lambda_{\mathrm{SO}}\sum_{\langle\langle ij \rangle\rangle \gamma \gamma'} \nu_{ij} c^{\dagger}_{i,\gamma} s^{z}_{\gamma,\gamma'} c_{j,\gamma'} + \nonumber \\
    & + i \lambda_{\mathrm{R}} \sum_{\langle ij\rangle\gamma \gamma'} \hat{z} \cdot (\vec{s}_{\gamma,\gamma'} \times \vec{d}) c^{\dagger}_{i,\gamma} c_{j,\gamma'}
\end{align}
where $\gamma$ denotes the spin, $E_p \xi_i$ is the onsite energy of the orbital $i$, $t$ is the hopping amplitude between the nearest neighbours. The third term on the right-hand side describes next-nearest neighbour spin-orbit coupling, $\vec{s}$ being the Pauli matrices and $\nu_{ij} = \pm 1$, depending on the orientation of the two nearest-neighbour bonds $\vec{d}_1$ and $\vec{d}_2$ along which the electron hops from orbital $i$ to orbital $j$. Finally, the fourth term describes nearest-neighbour Rashba spin-orbit coupling.

The Kane-Mele model describes a system on a honeycomb lattice with two sites (four orbitals) per unit cell. In the absence of the Rashba term, the spin is conserved. For two occupied orbitals in a unit cell the model describes an insulator if either $\lambda_{\mathrm{SO}}$ or $E_p$ is non-zero. For $\lambda_{\mathrm{SO}}=0$, the insulator is trivial and there is zero bulk SHC. If $E_p=0$ and $\lambda_{\mathrm{SO}}$ is finite, the Kane-Mele model becomes a topological insulator with a quantized bulk SHC $\sigma^{z}_{xy} = e/(2\pi)$~\cite{Kane2005a}. This quantization is violated by switching-on the Rashba spin-orbit coupling, which breaks the conservation of the spin.

\subsection{LSHC of an insulator}
\label{sec:LSHC_ins}
In the first test we choose the model parameters to be $E_p=0$, $t=1$, $\lambda_{\mathrm{SO}} = 0.1$, and $\lambda_{\mathrm{R}}=0$, modelling graphene with finite spin-orbit coupling, which is a $\mathcal{Z}_2$ topological insulator at half filling. The bulk SHC calculated with Eq.~\eqref{eq:shc3} on a  $50\times 50$ $\k$-point mesh yields $\sigma^{z}_{xy} = -e/(2\pi)$, as expected~\cite{Kane2005a}.

\begin{figure}
  \centering
  \includegraphics[width= 0.95\columnwidth]{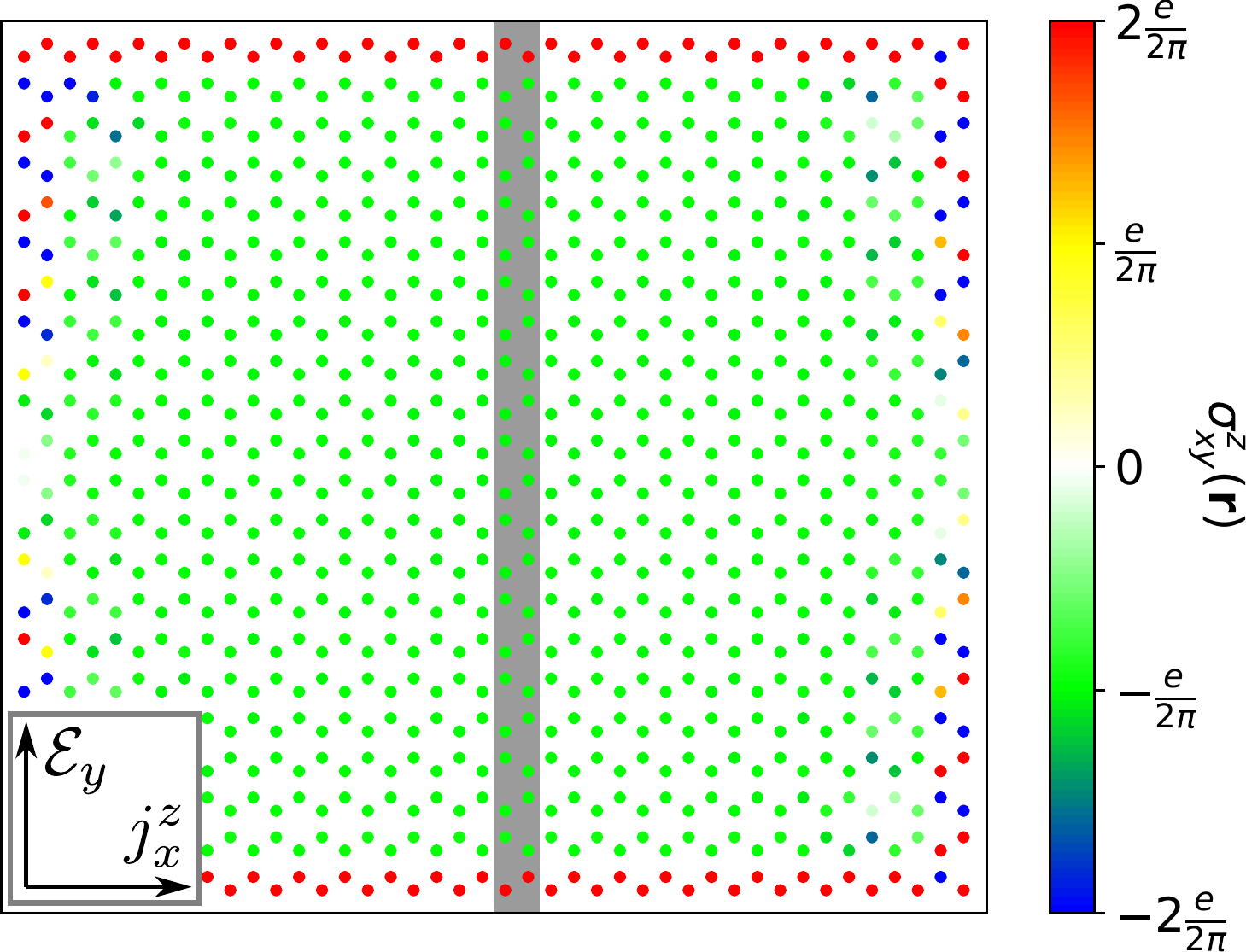}
  \caption{LSHC $\sigma^{z}_{xy}(\rdep)$ of a 924-sites flake of the Kane-Mele model in the topological insulating regime. Thick grey line refers to the values plotted in Fig.~\ref{fig:fig2}.}
  \label{fig:fig1}
\end{figure}

We then calculate the LSHC of the model using Eq.~\eqref{eq:lshc_fin}. For this purpose we construct a $1\times 2$ rectangular supercell and multiply it to create a large rectangular flake of the Kane-Mele system. The result is shown in Fig.~\ref{fig:fig1}.
The LSHC ammounts exactly to $-e/(2\pi)$ in the middle of the flake and it thus corresponds with the bulk value. The total SHC of the finite system is zero, since there cannot be any total currents in a system with OBC. This behaviour is analogous to the AHC discussed in Refs.~\onlinecite{Bianco2011,Marrazzo2017}. Therefore, the negative LSHC in the middle of the flake must be canceled by boundary contributions of opposite sign. The thickness of the boundary is $\sim$1 unit cell and the LSHC converges very fast to its bulk value. Indeed, exponential convergence towards the bulk value with the flake size is expected for insulators, owing to the exponential decay of the one-body density matrix~\cite{Kohn1996,Bianco2011}.

\begin{figure}
  \centering
  \includegraphics[width= 0.95\columnwidth]{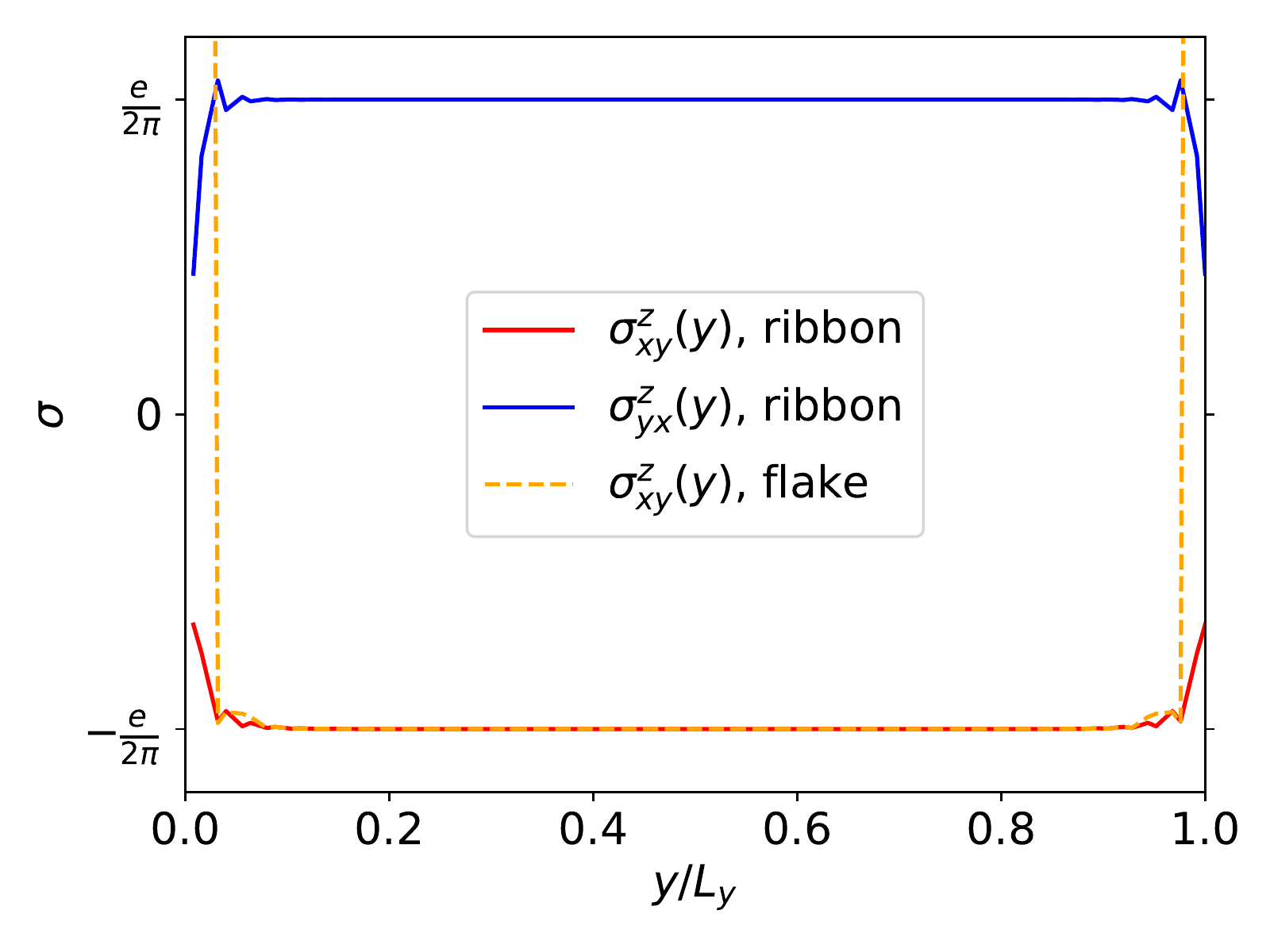}
  \caption{LSHC $\sigma^{z}_{xy}(y)$ (red) and $\sigma^{z}_{yx}(y)$ (blue) of a 84-sites thick nanoribbon (periodic along $x$, bounded along $y$) of the Kane-Mele model in the topological insulating regime. $L_y$ is the thickness of the nanoribbon. Orange dashed line shows $\sigma^{z}_{xy}(y)$ for sites in the grey region in Fig.~\ref{fig:fig1}.}
  \label{fig:fig2}
\end{figure}

In the next step we calculate the LSHC of a nanoribbon with PBC along $x$ and OBC along $y$ constructed multiplying the same $1\times 2$ supercell along the $y$-direction. The model parameters remained as in the previous calculation and the width of the nanoribbon was 21 supercells (84 sites). We calculated both $\sigma^{z}_{xy}(y)$ and $\sigma^{z}_{yx}(y)$ with the electric field oriented along $y$- and $x$-direction, using Eqs.~\eqref{eq:shc_a_eq_b} and \eqref{eq:shc_a_eq_a}, respectively. We show the results in Fig.~\ref{fig:fig2}.
Again, we obtained $\sigma^{z}_{xy}(y)=-e/(2\pi)$ in the middle of the nanoribbon, in perfect agreement with both the flake and bulk calculations. For the rotated geometry with electric field along the PBC direction we got $\sigma^{z}_{yx}(y)=e/(2\pi)$ in the bulk region, which also agrees with the previous results via $\sigma^{z}_{xy} = -\sigma^{z}_{yx}$. As for the finite system, the LSHC at the edges of the nanoribbon differs from the bulk region. Yet in this case the total SHC of the system does not vanish, since the system is periodic (and not finite) in the $x$-direction.

\subsection{LSHC of a metal}
\label{sec:LSHC_met}
In our second test we turn to a metallic system. We chose the model parameters $E_p=0$, $t=1$, $\lambda_{\mathrm{SO}} = 0.3$, and $\lambda_{\mathrm{R}}=0.25$ and set the Fermi level $E_{\mathrm{F}}=0.9$ to be located in the empty states. A bulk calculation with $200\times 200$ $\k$-points for this set of parameters yields $\sigma^{z}_{yx}=$ \unit[0.61]{e/($2\pi$)}.

As discussed in Ref.~\onlinecite{Marrazzo2017}, the one-body density matrix decays as a power law in metals, in contrast to the exponential decay in insulators. 
This leads to a more delocalized effect of inhomogenities on local properties, which can be for example Friedel oscillations or surface resonant states. These oscillations typically appear on a scale larger than a single bulk unit cell and can influence local material properties far from their origin.
We calculated the LSHC of a metallic nanoribbon constructed from up to 404 sites, performing the $\k$-integration along the periodic direction over 500 $\k$-points.
\begin{figure*}
  \centering
  \includegraphics[width= 0.66\columnwidth]{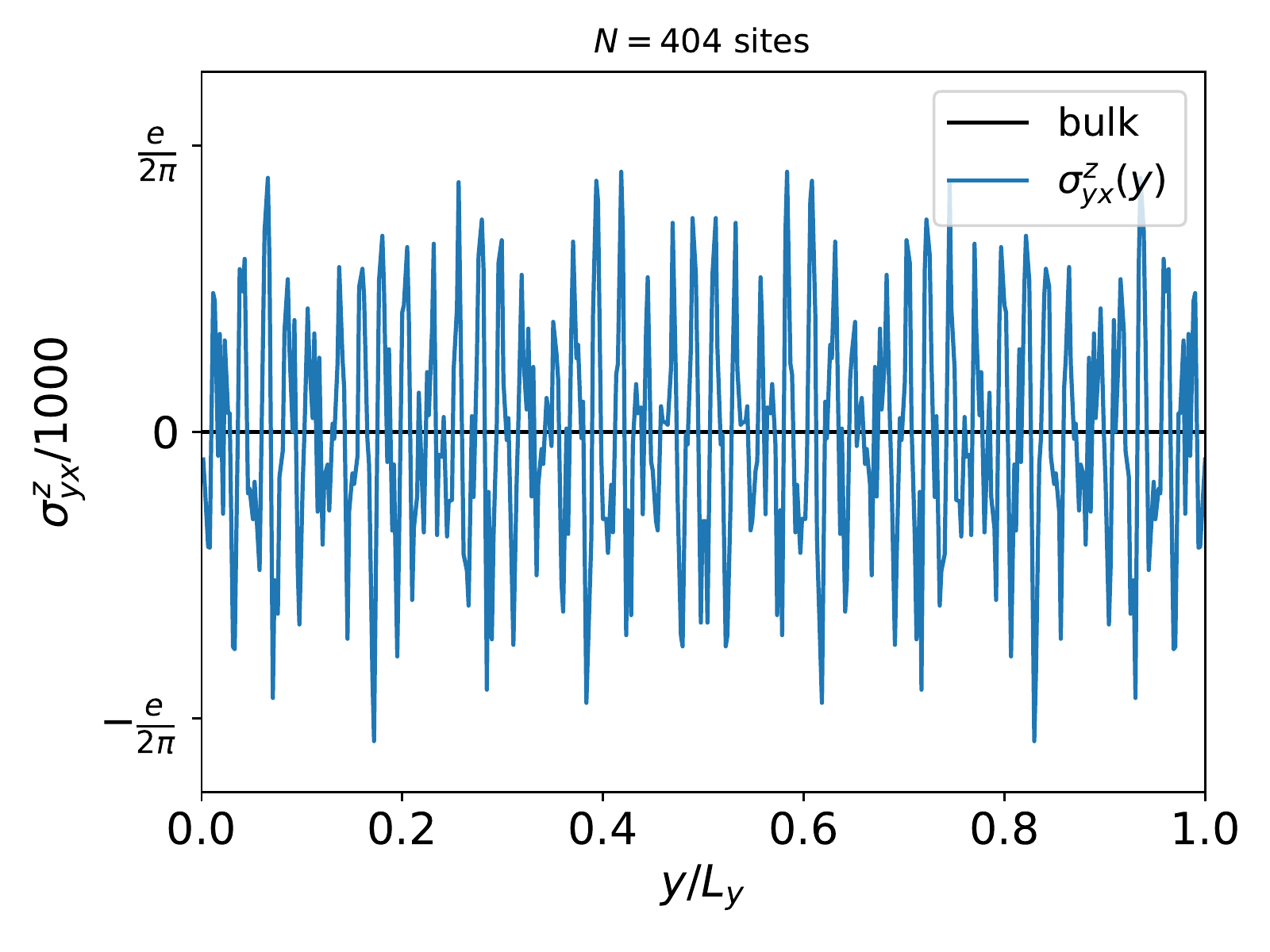}
  \includegraphics[width= 0.66\columnwidth]{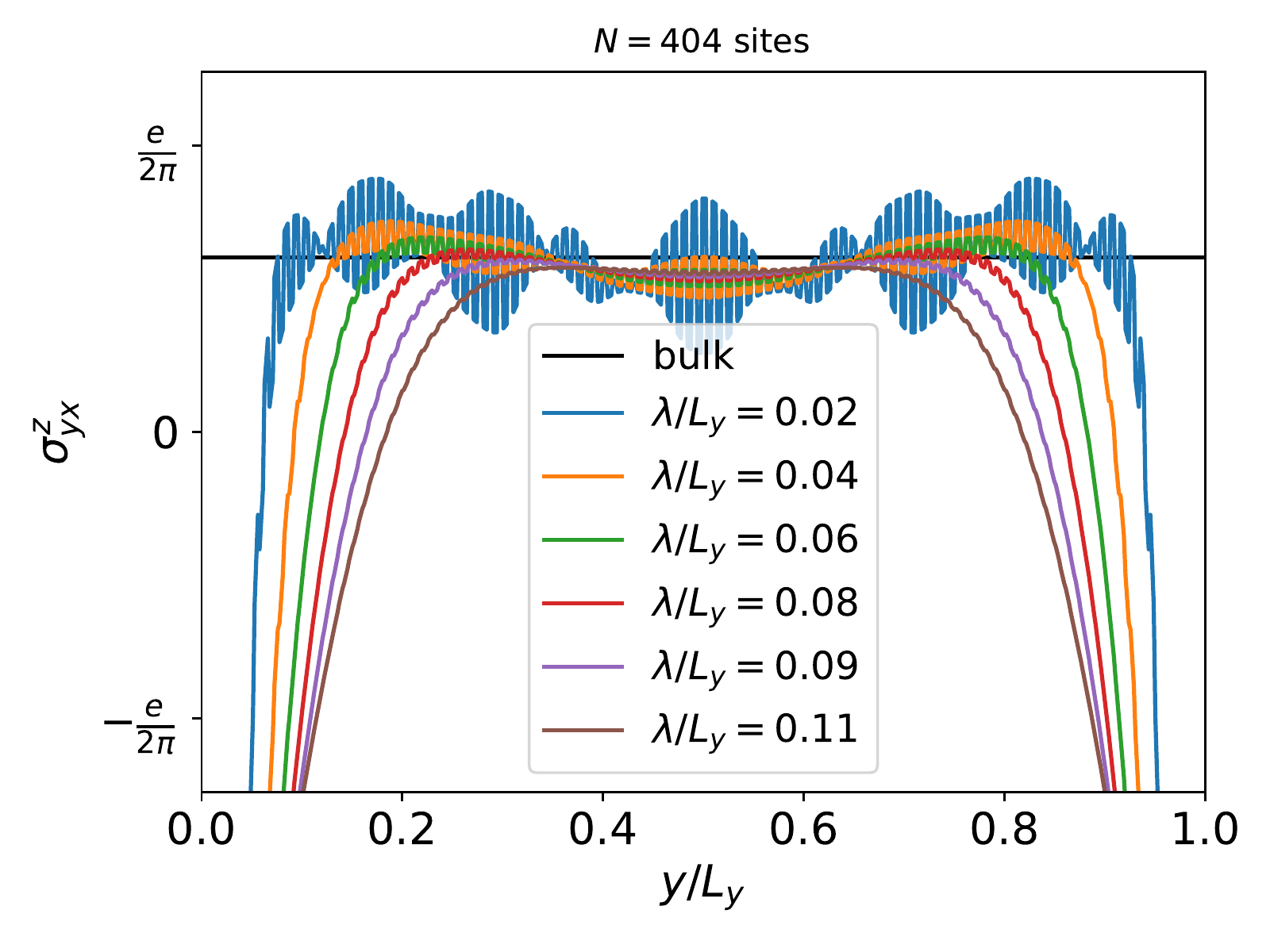}
  \includegraphics[width= 0.66\columnwidth]{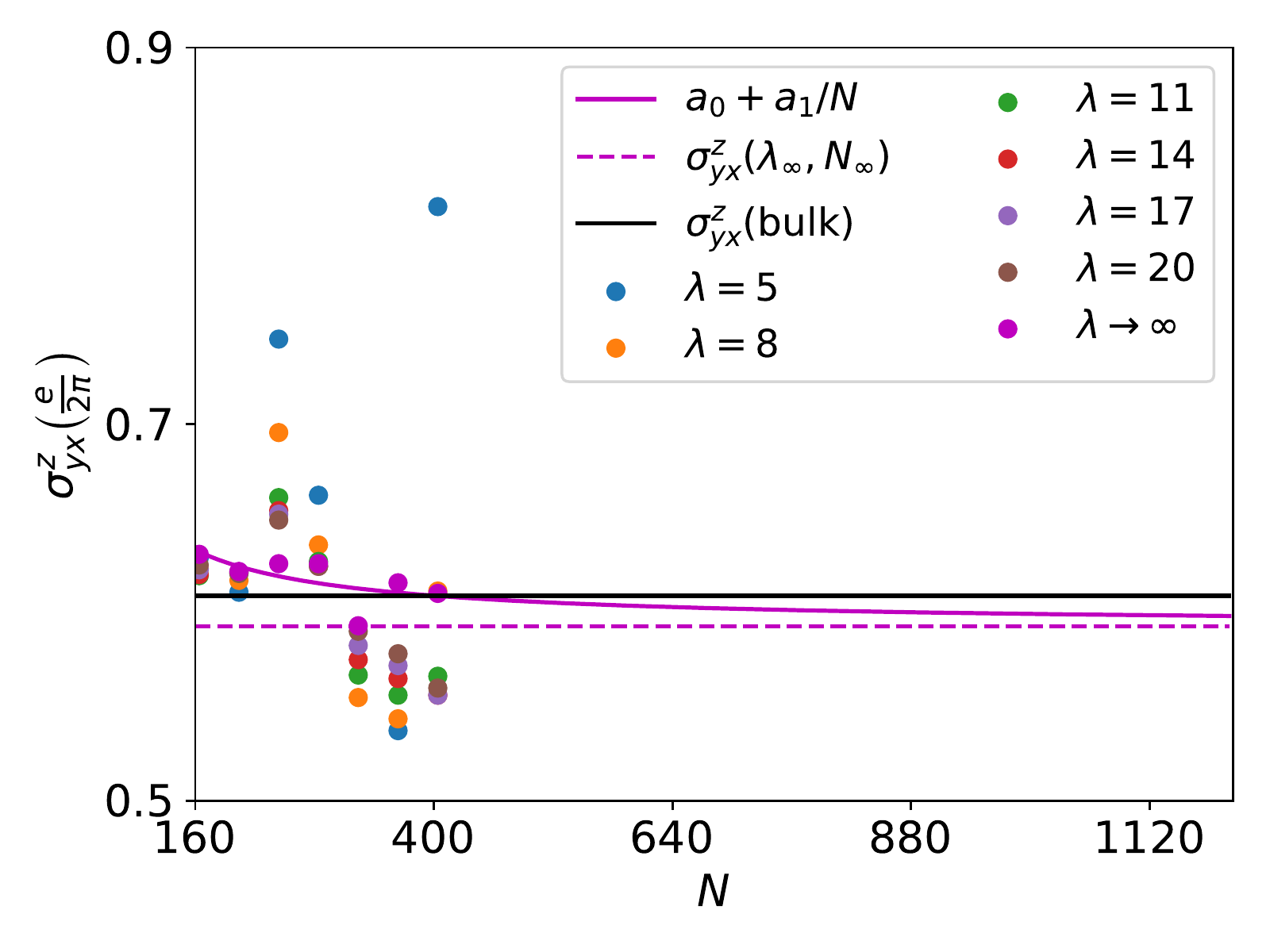}
  \caption{LSHC of a metallic nanoribbon. Left: Site-resolved LSHC for $N=404$ sites. Center: Gaussian-smeared site-resolved LSHC with different values of the smearing parameter $\lambda$. Right: Gaussian-smeared LSHC for different values of $\lambda$ and $N$ (dots), extrapolated to $\lambda\rightarrow \infty$ (magenta dots). The extrapolated values were fitted to $a_0 + a_1/N$ (magenta line), from which the $N\rightarrow \infty$ was extracted and plotted as dashed magenta line and compared with the bulk value (black line).}
  \label{fig:fig3}
\end{figure*}
We show the result for $N=404$ sites in the left panel of Fig.~\ref{fig:fig3}. The LSHC shows extreme oscillations of up to $\pm 1000 e/(2\pi)$ 
which can be of physical origin as mentioned above, as well as of numerical origin, caused by the finite number of $\k$-points used in a numerical calculation. In a realistic material, the physical oscillations are typically weakened by the averaging effect of temperature or disorder. In these test calculations we intent to demonstrate that the LSHC values in a locally homogeneous system agree with those calculated for a corresponding bulk periodic system. Therefore, we locally average the calculated LSHC values, effectively reducing both types of oscillations. To achieve this,
we apply a Gaussian smearing to the LSHC at site $l$ as
\begin{equation}
    \label{eq:lshc_smeared}
    \overline{\sigma}^{z}_{yx}(l,\lambda) = \frac{\sum_{l'}{ \sigma^{z}_{yx}(l')\  e^{-\frac{(\r_l-\r_{l'})^2}{2\lambda^{2}}}}}{\sum_{l'}{e^{-\frac{(\r_l-\r_{l'})^2}{2\lambda^{2}}}}}.
\end{equation}
Depending on the parameter $\lambda$, the LSHC at each site is averaged over a certain neighboring region.
We show in the central panel of Fig.~\ref{fig:fig3} that with increasing $\lambda$ the locally averaged LSHC approaches the bulk SHC value in the center of the nanoribbon. Assuming an infinitely large nanoribbon ($N\rightarrow \infty$) and a global average ($\lambda \rightarrow \infty$), we expect to obtain the bulk value exactly. Therefore, we performed a set of calculations for $N\in (164,404)$ and $\lambda \in (5,20)$. In the right panel of Fig.~\ref{fig:fig3} we plot the values of the smeared LSHC $\overline{\sigma}^{z}_{yx}(l_{c},\lambda,N)$ for a site $l_c$ in the center of the nanoribbon. For each $N$ we fitted the values to $\overline{\sigma}^{z}_{yx}(l_{c},\lambda,N)\propto \lambda^{-2}$ and obtained $\overline{\sigma}^{z}_{yx}(l_{c},\lambda_{\infty},N)$, which we then fitted to $\overline{\sigma}^{z}_{yx}(l_{c},\lambda_{\infty},N)\propto N^{-1}$ and extrapolated for $N \rightarrow \infty$. In the right panel of Fig.~\ref{fig:fig3} we show the fitted LSHC $\overline{\sigma}^{z}_{yx}(l_{c},\lambda_{\infty},N) = a_0 + a_1/N$ as a full magenta line and the extrapolated value $\overline{\sigma}^{z}_{yx}(l_{c},\lambda_{\infty},N_{\infty})=$ \unit[0.59]{e/($2\pi$)}. This value shows a relative error of $\sim 3\%$ with respect to the bulk value, which confirms the validity of the LSHC formulas also for metals.

\subsection{LSHC of heterostructures}
\begin{figure}
  \centering
  \includegraphics[width= 0.99\columnwidth]{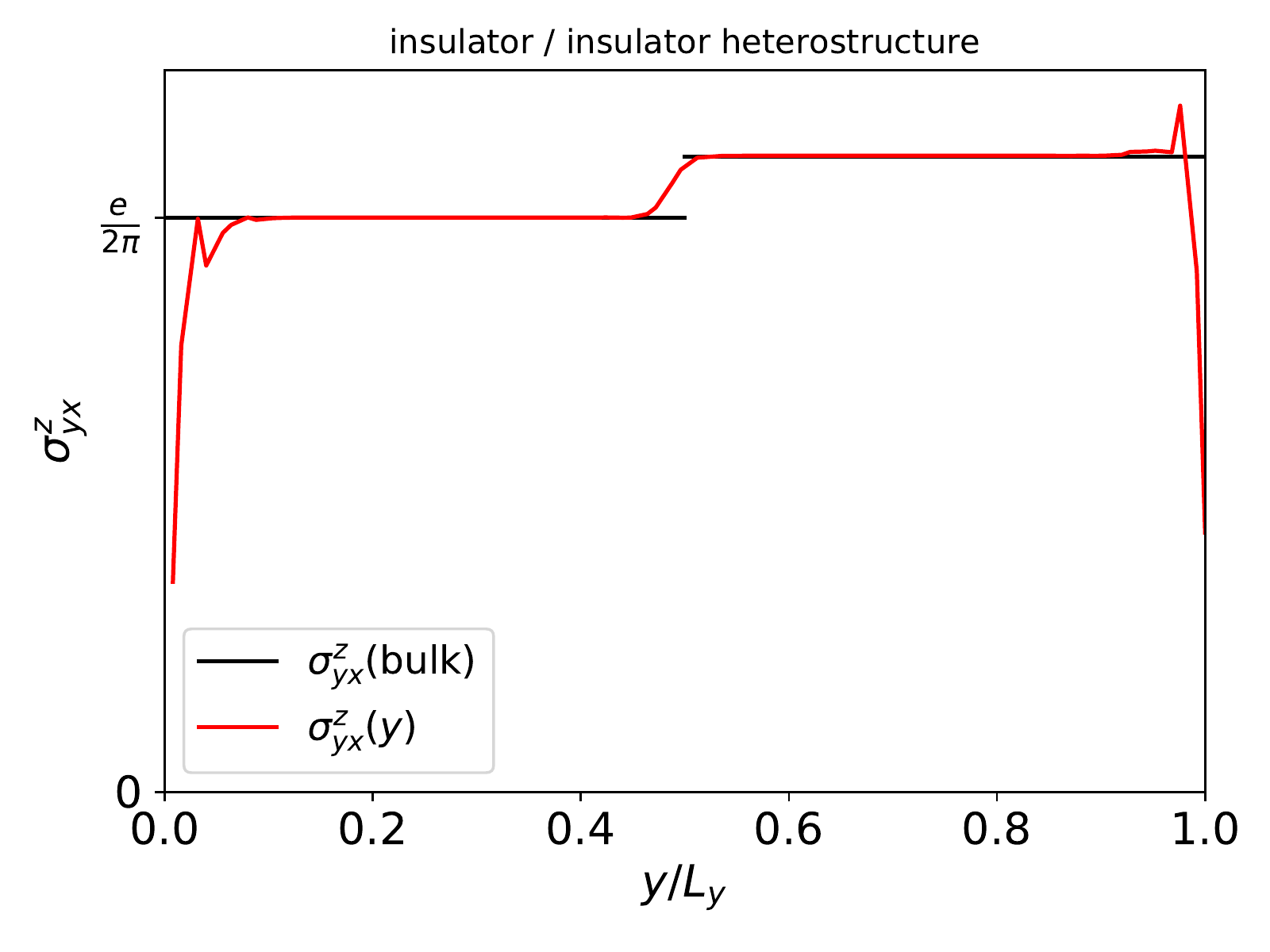}\\
  \includegraphics[width= 0.99\columnwidth]{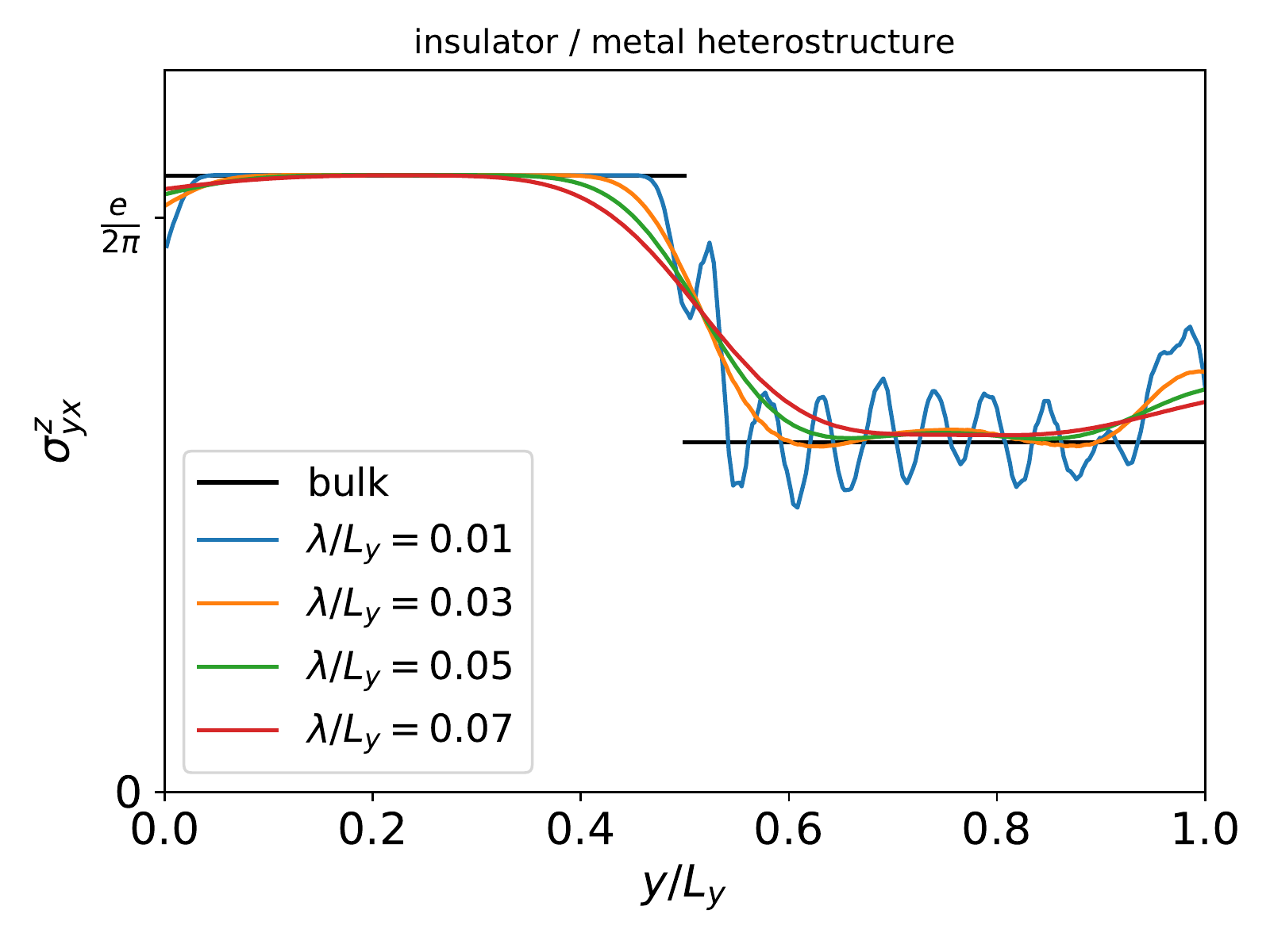}\\
  \caption{LSHC of a heterostructure. Top: Insulator / insulator heterostructure, original LSHC values. Bottom: Insulator / metal heterostructure, averaged LSHC values with different smearing $\lambda$.}
  \label{fig:fig4}
\end{figure}

Finally, we test our local approach on a nanoribbon composed of two different parts. First, we construct an insulator / insulator heterostructure, with one part being described by the Kane-Mele Hamiltonian with the parameters $E_p=0$, $t=1$, $\lambda_{\mathrm{SO}} = 0.3$, and $\lambda_{\mathrm{R}}=0.0$, which has a bulk SHC $\sigma^{z}_{yx}=$ e/($2\pi$). The Hamiltonian of the second part is specified by the parameters $E_p=0$, $t=1$, $\lambda_{\mathrm{SO}} = 0.3$, and $\lambda_{\mathrm{R}}=0.3$. The bulk SHC is $\sigma^{z}_{yx}=$ \unit[1.106]{e/($2\pi$)}, owing to the nonzero Rashba-term which breaks the spin conservation. The results of the calculation for the nanoribbon geometry ($N=84$ sites, $50\ \k$-points) are shown in the top panel of Fig.~\ref{fig:fig4}.
The LSHC in the center of each constituent equals the respective bulk SHC values. Averaging the LSHC is not necessary, because of the exponential decay of the ground-state projector, as described in Sec.~\ref{sec:LSHC_ins}.

As a second heterostructure we chose an insulator / metal system. The parameters of the insulating part were $E_p=0$, $t=1$, $\lambda_{\mathrm{SO}} = 0.3$, and $\lambda_{\mathrm{R}}=0.25$ with a bulk SHC $\sigma^{z}_{yx}=$ \unit[1.073]{e/($2\pi$)}. The metallic part was characterized by the same set of parameters, but the orbitals were shifted energetically down by $0.9$, so $E_{\rm F}$ is located in the conduction bands. The subsystem equals the one in Sec.~\ref{sec:LSHC_met} and the bulk SHC is $\sigma^{z}_{yx}=$ \unit[0.6088]{e/($2\pi$)}. The nanoribbon was constructed from $N=324$ sites and we used $500\ \k$-points for the integration in the reciprocal space. Since the LSHC values are again strongly oscillating in the metallic region, we applied the averaging given by Eq.~\eqref{eq:lshc_smeared}. We present the results in the bottom panel of Fig.~\ref{fig:fig4}, which show a perfect agreement with the bulk SHC for all values of $\lambda$ in the insulating part. Following the same extrapolation procedure of the averaged LSHC value in the middle of the metallic region for $\lambda \rightarrow \infty$, we obtain $\sigma^{z}_{yx}=$ \unit[0.6107]{e/($2\pi$)}, which has a relative error of $0.3\%$ with respect to the corresponding bulk value.

We thus state that our numerical model calculations fully support the analytical formulas for the LSHC with both OBC and mixed boundary conditions. For insulating systems, the expected bulk SHC value is recovered close to the heterogeneous regions, whereas for metallic systems the convergence with system size is much slower ($\sim 1/N$ for the nanoribbon) and averaging over a sufficiently large region (i.e., choosing large $\lambda$) is necessary. On the other hand, 
for practical uses,
$\lambda$ has to be kept small enough to capture the local deviations of the LSHC from the bulk value in inhomogeneous regions.

\section{Conclusion}
\label{sec:conclusion}
Led by the analogies between the anomalous and spin Hall conductivity, we have derived an expression for a local SHC in two-dimensional materials, starting from the local spin current density. Working in the framework of a bounded finite system, we further checked that performing the thermodynamic limit and assuming an infinitely large unbounded periodic system, our LSHC expression leads to the usual Kubo-formula for SHC. Furthermore, assuming PBC in one direction and keeping the system bounded with OBC in another direction, we derived an expression for the LSHC in the nanoribbon geometry, which is possibly computationally faster and easier parallelizable than the fully bounded system.

In the next step we have applied our analytical results to the Kane-Mele model, which has a finite bulk SHC in both the metallic and insulating regime, owing to its non-trivial topological character. We have shown that for both regimes the LSHC in the central region of a large nanoribbon or flake equals the SHC of a referent bulk system. Whereas the equality of both approaches is achieved rapidly already for small systems in the case of insulators, for metals an averaging procedure over a sufficiently large region and extrapolation to an infinitely large system is necessary. For finite inhomogeneous metallic systems, one thus has to choose a compromise between averaging over a sufficiently large region to suppress the strong oscillations on one side, and averaging over a small enough region to keep the local information.

We expect our results to bring new insight into spintronics phenomena of inhomogeneous systems, such as heterostructures, surfaces, defects, etc. In particular, we propose to use the LSHC to calculate the influence of system inhomogenities to the spin Hall effect, which we assume to alter the bulk values significantly.

Note that first attempt to carry the formalism of the local Chern marker to the context of spin Hall insulators can be found in Ref.~\onlinecite{Amaricci2017}, where the ``local $\mathbb{Z}_2$ invariant'' has been calculated as the spin Chern number, under the assumption that the system Hamiltonian can be separated into distinct spin-up and spin-down subspaces. The original local Chern marker can then be calculated as introduced in Ref.~\onlinecite{Bianco2011} for each subspace individually, to give the local spin Chern number.

\begin{acknowledgments}
This work was supported by the CRC TRR 227 of Deutsche Forschungsgemeinschaft (DFG) and by the Forschungsstipendium Grant No. RA 3025/1-1 from the DFG (T.R.). We thank Ivo Souza for stimulating discussions.
\end{acknowledgments}

\bibliographystyle{apsrev4-1}
\bibliography{paper.bbl}

\end{document}